\documentclass[twocolumn,amssymb,amsmath,aps,prb,showpacs]{revtex4-2}
\usepackage{array}
\usepackage{graphicx}
\usepackage{color}
\usepackage{multirow}
\usepackage{float}
\usepackage{amssymb}
\usepackage{amsmath}

\begin{document}

\title{Successive ferroelectric orders and magnetoelectric coupling without long-range magnetic order in highly frustrated pyrochlore compounds: Sm$_2$Ti$_{2-x}$V$_x$O$_7$}
%\title{Exploring Ferroelectric and Magnetic Properties in Frustrated Pyrochlore Compounds: Insights from Sm$_2$Ti$_2$O$_7$ and Vanadium-Doped Variant}
%\title{Enhancing magnetoelectric coupling through V-doping below ferroelectric order in frustrated pyrochlore compounds: Sm$_2$Ti$_{2-x}$V$_x$O$_7$}
\author{S. Mukherjee$^1$}
\author{O. Ivashko$^2$}
\author{S. Majumdar$^1$}
\author{A. Kumar$^3$}
\author{S. Giri$^{1}$}
\email{Corresponding author: sspsg2@iacs.res.in} 
\affiliation{$^1$School of Physical Sciences, Indian Association for the Cultivation of Science, Jadavpur, Kolkata 700032, India}
\affiliation{$^2$Deutsches Elektronen-Synchrotron DESYNotkestr. 85, 22607 Hamburg, Germany}
\affiliation{$^3$Solid State Physics Division, Bhabha Atomic Research Centre, Mumbai 400085, India
and Homi Bhabha National Institute, Anushaktinagar, Mumbai 400094, India}
%\begin{document}
\begin{abstract}
Sm$_2$Ti$_2$O$_7$, a member of rare-earth titanate pyrochlores, exhibits dipolar-octupolar antiferromagnetism below $T_N$ = 0.35 K. We observed two ferroelectric transitions at 182 ($T_{FE1}$) and 52 K ($T_{FE2}$), significantly higher than $T_N$ for Sm$_2$Ti$_{2-x}$V$_x$O$_7$ ($x$ = 0, 0.1). Although the  ferroelectric transition temperatures remain unchanged, the polarization value decreases considerably with V doping. A structural transition to a polar $R3m$ rhombohedral phase from the cubic $Fd\bar{3}m$ structure occurs at $T_{FE1}$, involving a distortion in the pyochlore lattice. Remarkably, significant linear magnetoelectric coupling is observed in both compounds, with further enhancement of magnetoelectric coupling due to magnetic V  doping. The existence of magnetoelectric coupling without long-range magnetic order in a frustrated pyrochlore system could enable the tailoring of magnetoelectric coupling properties, which can be further fine-tuned through V doping. The emergence of ferroelectricity in a frustrated magnetic system introduces an intriguing aspect to these compounds and paves the way for developing ferroelectric order driven by the alleviation of magnetic frustration in pyrochlore systems.

\end{abstract}
%
%\pacs{75.85.+t, 77.80.-e, 75.80.+q}
%75.85.+t, 77.80.-e
\maketitle

\section{Introduction}
The frustrated pyrochlore structure, manifesting in compounds with the chemical formula $R_2$B$_2$O$_7$ \cite{Moess} ($R$ = Rare earth elements; M=Zr, Hf, Ti, Sn, Pb), stands as a captivating archetype in the realm of condensed matter physics and materials science. This structure, defined by a lattice of corner-sharing tetrahedra formed by $R$ cations interconnected by oxygen ions, embodies a symphony of competing interactions and geometric frustrations.
The delicate interplay between the magnetic frustration inherent in the pyrochlore lattice and the magnetic exchange interactions often gives rise to a plethora of exotic physical phenomena, spanning from unconventional magnetic ground states to intriguing electronic behaviors.
Within the domain of magnetism, frustrated pyrochlores have garnered significant attention due to their propensity to host intriguing magnetic phases such as spin liquids \cite{moes,ben}, spin ice \cite{Bram,Bovo,Ramirez1}, and spin glasses \cite{shin,mit}. 

The pyrochlore family, represented by the chemical formula $R_2$Ti$_2$O$_7$ (where $R$ ranges from Gd to Yb), is renowned for being exemplary three-dimensional geometrically frustrated systems \cite{Greedan}. This recognition stems from their intriguing and unconventional magnetic characteristics. Within these insulating pyrochlore structure, a variety of rich characteristic features have been witnessed. These include suppressed moment in Gd$_2$Ti$_2$O$_7$ \cite{padd,pet}, single-ion anisotropy of Tb$^{3+}$ within the collective paramagnetic-spin liquid state in Gd$_2$Ti$_2$O$_7$ \cite{Gingras,Gardner}, and the realization of quantum order by disorder and accidental soft mode in Er$_2$Ti$_2$O$_7$ \cite{zhit,bri}. On the other hand, intriguing magnetic behavior from spin ice state to supercooled spin liquid state has been observed in Dy$_2$Ti$_2$O$_7$ \cite{sam,kas}, the emergence of magneic monopoles has been noted in Gd$_2$Ti$_2$O$_7$ \cite{fenn,bor,Harris}, and evidence of quantum spin-liquid behavior has been found in Yb$_2$Ti$_2$O$_7$ \cite{kerm,sch}.

%%%%%%%%%%
\begin{figure}[b]
\centering
\includegraphics[width = \columnwidth]{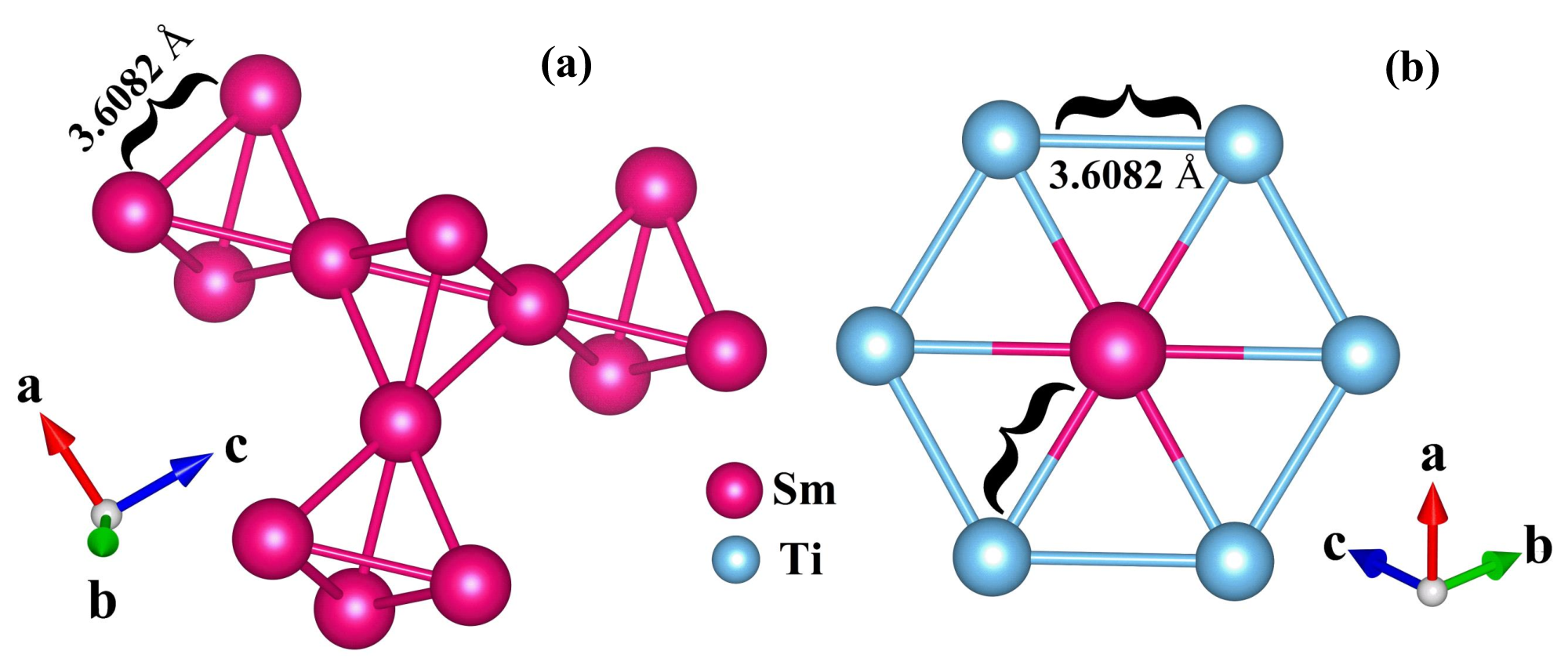}
\caption {(a) Equidistant pyrochlore structure of STO formed by Sm atoms and (b) formation of an equidistant hexagonal structure by Ti atoms, with Sm atoms sitting at the center.}
\label{structure}
\end{figure}
%%%%%%%%%%%%%%%

Magnetism of our focal compound, Sm$_2$Ti$_2$O$_7$, as a member of pyrochlore $R_2$Ti$_2$O$_7$ family, has been relatively underexplored at the microscopic level, particularly through neutron scattering. This is due to the fact that $^{149}$Sm is a strong neutron absorber.  Nevertheless, recent neutron scattering, facilitated by enriching the $^{154}$Sm isotope, and muon relaxation studies have revealed that the compound exhibits the all-in, all-out magnetic structure with an ordered moment of 0.44(7)$\mu_B$ below $T_N$ = 0.35 K. This observation aligns with expectations for antiferromagnetically coupled Ising spins on the pyrochlore lattice \cite{mau,bent}. Consistent with this findings, bulk magnetization studies do not indicate long range magnetic order down to 0.5 K. This absence was attributed to weak exchange (Curie-Weiss temperature $\sim$ -0.26 K) and dipolar interactions ($\mu_{eff} \sim$ 0.15 $mu_B$) between the Sm$^{3+}$ spins within this pyrochlore structure \cite{sing1}. 

%Consistent with the previous reports, the Rietveld refinement of our synchrotron X-ray diffraction satisfactorily fits the diffraction pattern at 300 K with $Fd\bar{3}m$ space group, as shown in Fig. \ref{structure}(a) \cite{FARMER}. The Rietveld refinement offers insights into the local structures, as illustrated in Figs. \ref{structure}(b) and \ref{structure}(c). Figure \ref{structure}(b) portrays regular pyrochlore structure, where all vertices of a regular tetrahedron are equidistant from each other, measuring a length of 2.6082 $\AA$ at 300 K. Figure \ref{structure}(c) showcases the equidistant (2.6082 $\AA$) hexagonal structure formed by Ti atoms, with magnetic Sm$^{3+}$ occupying the center. Incidentally, Ti atoms form edge-sharing equilateral triangles, with the Sm atom situated at one of the vertices. Now, if we replace the Ti$^{3+}$ ion with magnetic V$^{4+}$, the magnetic V$^{4+}$ ions, arranged in an equilateral triangle, influence frustrated magnetism due to pyrochlore structure formed by Sm$^{3+}$ ions and act as a perturbation to the frustrated magnetism of Sm$_2$Ti$_2$O$_7$. 

\begin{figure}[h!]
\centering
\includegraphics[width = \columnwidth]{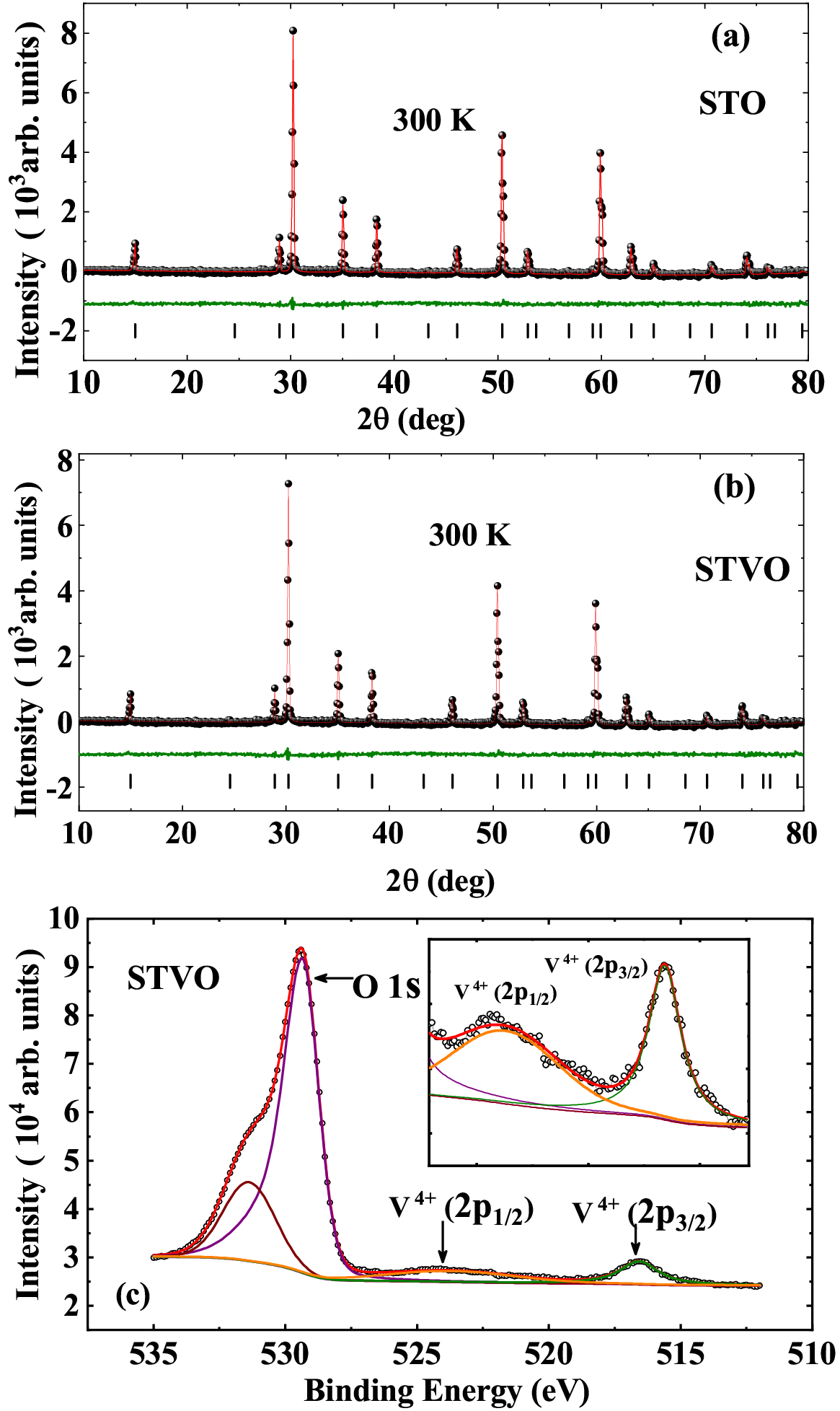}
\caption { Rietveld refined XRD patterns at 300 K of (a) STO and (b) STVO. (c) XPS spectrum of STVO. Inset shows deconvolution of (solid curve) of 2$p_{1/2}$ and 2$p_{3/2}$ spectrum of V$^{4+}$ for STVO. }
\label{XRD}
\end{figure}

Recent developments suggest the potential occurrence of ferroelectric (FE) order in the pyrochlore $R_2$Ti$_2$O$_7$ family, where $R$ = La, Ce, Pr, and Nd, based on first-principles studies utilizing density functional theory \cite{spaldin}. While the search for FE order has rarely been attempted in polycrystalline materials \cite{Patwe,NANAMATSU} and films \cite{Shao,Saitzek}, we are motivated to explore FE order as well as magnetoelectric coupling in Sm$_2$Ti$_{2-x}$V$_x$O$_7$ ($x$ = 0, 0.1), members of this pyrochlore family. In line with previous findings, Fig. \ref{structure}(a) depicts the characteristic pyrochlore structure of Sm$_2$Ti$_2$O$_7$, wherein all vertices of a regular tetrahedron are equidistant from each other, measuring a length of 3.6082 $\AA$ at 300 K, as obtained from Rietveld refinement (Fig. \ref{XRD}(a)). Figure \ref{structure}(b) illustrates  the formation of an equidistant (3.6082 $\AA$) hexagonal structure by Ti atoms, with magnetic Sm$^{3+}$ occupying the center. Notably, Ti atoms form Edge-sharing equilateral triangles, with the Sm atom positioned at the common vertices. Now, if we replace the Ti$^{4+}$ ion with magnetic V$^{4+}$, the magnetic V$^{4+}$ ions, arranged in an equilateral triangle, influence frustrated magnetism due to the pyrochlore structure formed by Sm$^{3+}$ ions and act as a perturbation to the frustrated magnetism of Sm$_2$Ti$_2$O$_7$.
   
In this article, we present findings on FE order accompanied by a structural transition to a polar $R3m$ structure from the cubic $Fd\bar{3}m$ structure. Of particular significance is the observation of magnetoelectric (ME) coupling in the absence of long-range magnetic order, a phenomenon further amplified by V doping. We discuss the influence of structural distortion and potential short-range effects driven by V doping on the ME coupling.

%%%%%%%%%%%%%%%%%%%%%%%%%%%%%%%%%%%%%%%%%%%%%%%%
\section{Experimental details}
%%%%%%%%%%%%%%%%%%%%%%%%%%%%%%%%%%%%%%%%%%%%%%%%

%%%%%%%%%%%%%%%%%%%%%%%%%%%%%%%%%
%%%%%%%%%%%%%%%%%%

Polycrystalline Sm$_2$Ti$_2$O$_7$ (STO) and Sm$_2$Ti$_{1.9}$V$_{0.1}$O$_7$ (STVO) were synthesized using a standard solid-state reaction technique. Stoichiometric amounts of Sm$_2$O$_3$ and TiO$_2$ and V$_2$O$_5$ (for STVO) were finely ground and thoroughly mixed for several hours. This mixture was  first preheated at 500$^\circ$C for 5 hours. Subsequently, the mixtures were ground and thoroughly mixed before undergoing a final heat treatment at 1300$^\circ$C for 24 hours after being pressed into pellets \cite{tabira}.

The single-phase chemical compositions were verified using standard X-ray diffraction (XRD) studies at 300 K using a PANalytical X-ray diffractometer (Model: X' Pert PRO) with Cu-K$\alpha$ radiation. Low-temperature synchrotron X-ray diffraction studies were carried out in the range of 10-300 K using a wavelength, $\lambda$ = 0.1123 $\AA$ at P21.1 beam line at DESY, Hamburg, Germany. The diffraction patterns were analyzed using the Rietveld refinement  with a commercially available open-source software MAUD (Material Analysis Using Diffraction). 
Properly prepared pelletized samples were used for the pyroelectric measurements, electrical contacts made using silver paint. Pyroelectric current ($I_p$) was recorded in constant temperature sweep and magnetic field sweep rates using an electrometer (Keithley, model 6517B) connected to a commercial PPMS-II, EverCool system of Quantum Design. Poling electric fields were applied during the cooling process, and at the lowest temperature, the pellet was properly short-circuited before measuring $I_p$ during the heating process. Similarly, dielectric measurements were performed using an Agilent E4890A LCR meter connected with the same PPMS-II. Magnetization was measured using a commercial magnetometer of Quantum Design (MPMS, EverCool). X-ray photoemission spectroscopy (XPS) was recorded with a spectrometer of Omicron Nanotechnology.
%%%%%%%%%%%%%%%%%%%%%%%%%%%%%%%%%%%%%%%%%%%%%%%%

\section{Experimental results and discussions}

%\subsection{ Room temperature X-ray diffraction}
%%%%%%%%%%%%%%%%%%%%%%%%%%%%%%%%%%%%%%%%%%%%%%%

The XRD patterns at 300 K of STO and STVO are depicted in Figs. \ref{XRD}(a) and \ref{XRD}(b), respectively. These compounds crystallize in $Fd\bar{3}m$ space group, and Rietveld calculated XRD patterns are represented by the continuous curves in the figures. The goodness of fits for STO and STVO is satisfactory, confirming the absence of any impurity phase, as indicated by the difference plotted below. The Wyckoff positions for STO are as follows: Sm(0.5000, 0.5000, 0.5000), Ti(0, 0, 0), O1(0.3270(6), 0.1250, 0.1250), O2(0.3750, 0.3750, 0.3750), with a lattice constant, $a$ = 10.2815(3) $\AA$. The reliability parameters are $R_{w}$(\%) $\sim$ 3.47, $R_{exp}$(\%) $\sim$ 1.41, $\chi^2 \sim$ 1.56. The lattice constant at 300 K is close to the reported value for STO \cite{FARMER}. Here, all the Wyckoff positions are fixed except for $x$ for O1. The refined $x$ coordinate for STVO is 0.3272(6). For STVO, the refined lattice constant is $a$ = 10.2785(6) $\AA$, with  reliability parameters $R_{w}$(\%) $\sim$ 3.04, $R_{exp}$(\%) $\sim$ 1.78, $\chi^2 \sim$ 1.71. In line with the smaller crystal radius of six-coordinated V$^{4+}$ compared to Ti$^{4+}$, the lattice constant contracts upon V doping.  
To identify the oxidation state of V, XPS was recorded for STVO, as shown in Fig. \ref{XRD}(c). Satisfactory fits of the peaks are demonstrated by the continuous curves in the figure. The two peaks located at $524$ and $516$ eV indicate $V^{4+}$ states of $2p_{1/2}$ and $2p_{3/2}$, respectively, consistent with a previous report \cite{rakthe}, confirming the presence of V$^{4+}$ replacing Ti$^{4+}$.
%%%%%%%%%%%%%%%%%%%%%%%%%%%%%%%%%%%%%%%%%%%%%%%%
%\subsection{ DC magnetization}
%%%%%%%%%%%%%%%%%%%%%%%%%%%%%%%%%%%%%%%%%%%%%%%%

%%%%%%%%%%%%%%%%%%%%%%%%%%%%%%%%%%%%%%%%%%%%%%%%%%%%%%%%%%%%%%%%%%%%
\begin{figure}
\centering
\includegraphics[width = \columnwidth]{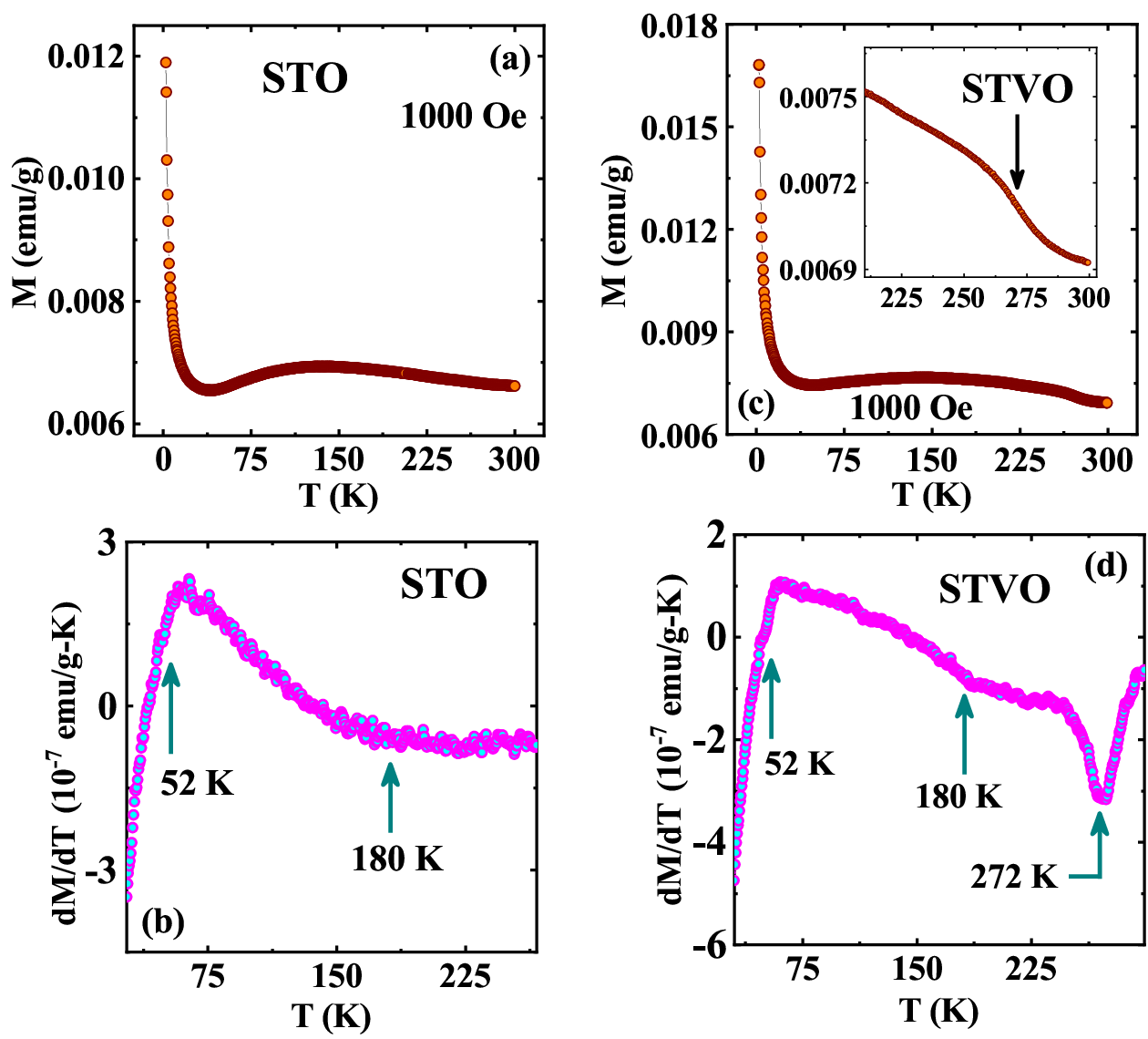}
\caption {(a) Thermal ($T$) variations of magnetization ($M$) recorded with 1000 Oe for (a) STO and (c) STVO. $T$ variations of $dM/dT$ for (b) STO and (d) STVO. Inset of (c) highlights an anomaly, as indicated by an arrow.}
\label{magnetic}
\end{figure}
%%%%%%%%%%%%%%%%%%%%%%%%%%%%%%%%%%%%%%%%%%%%%%%%
%%%%%%%%%%%%%%%%%%%%%%%%%
\begin{figure}[t]
\centering
\includegraphics[width = \columnwidth]{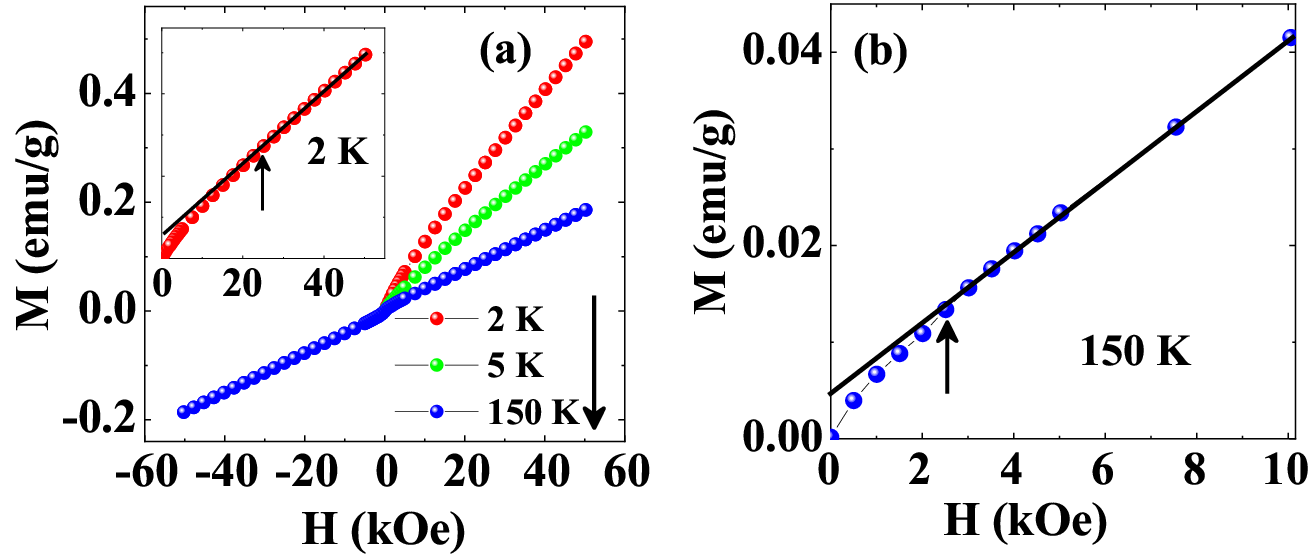}
\caption {(a) Isothermal magnetization of STVO at selected $T$. Inset of (a), and (b) show linear dependence of magnetic field ($H$) in the high-$H$ regions at 2 and 150 K, respectively.} 
\label{MH}
\end{figure}

%%%%%%%%%%%%%%%%%%%%%%

%Specifically, a sharp minimum is additionally observed at 272 K for STVO. The isothermal magnetization curves of STVO are depicted in Fig. 4(a) at selected T. Unlike the linear behavior of the magnetization curves for STO [24], non-linearity in the magnetization curves is evident at low H for STVO, as highlighted in the inset of Fig. 4(a) at 2 K and Fig. 4(b) at 150 K, respectively. Arrows in the figures indicate the magnetic field, beyond which linearity of the magnetization curves is observed, decreasing with increasing T. This non-linearity may correlate with the emergence of short-range magnetic order below 270 K for STVO.

%%%%%%%%%%%%%%%%%%%%%%%%%
\begin{figure}[b]
\centering
\includegraphics[width = 0.8\columnwidth]{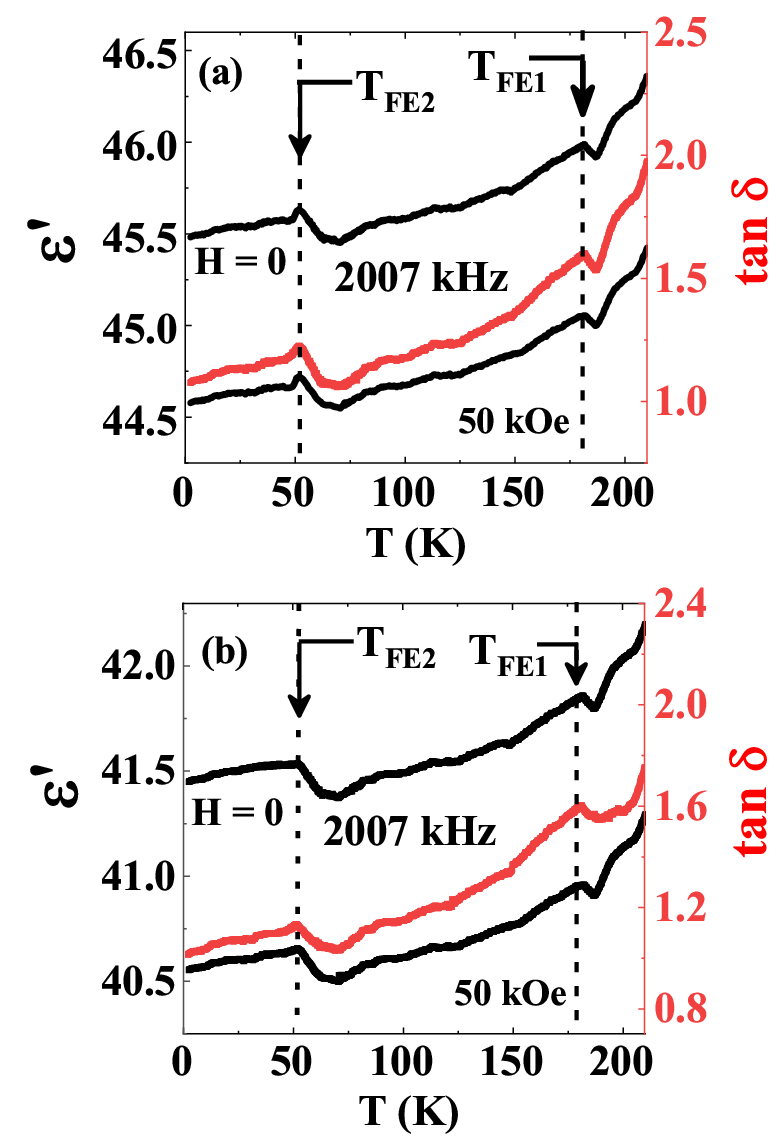}
\caption {$T$ variations of real part of dielectric constant ($\epsilon'$) at $H$ = 0 and 50 kOe, and dielectric loss ($\tan\delta$) at $H$ = 0 for (a) STO and (b) STVO, respectively.} 
\label{die}
\end{figure}
%%%%%%%%%%%%%%%%%%%%%%

%%%%%%%%%%
\begin{figure*}
\centering
\includegraphics[width = 2\columnwidth]{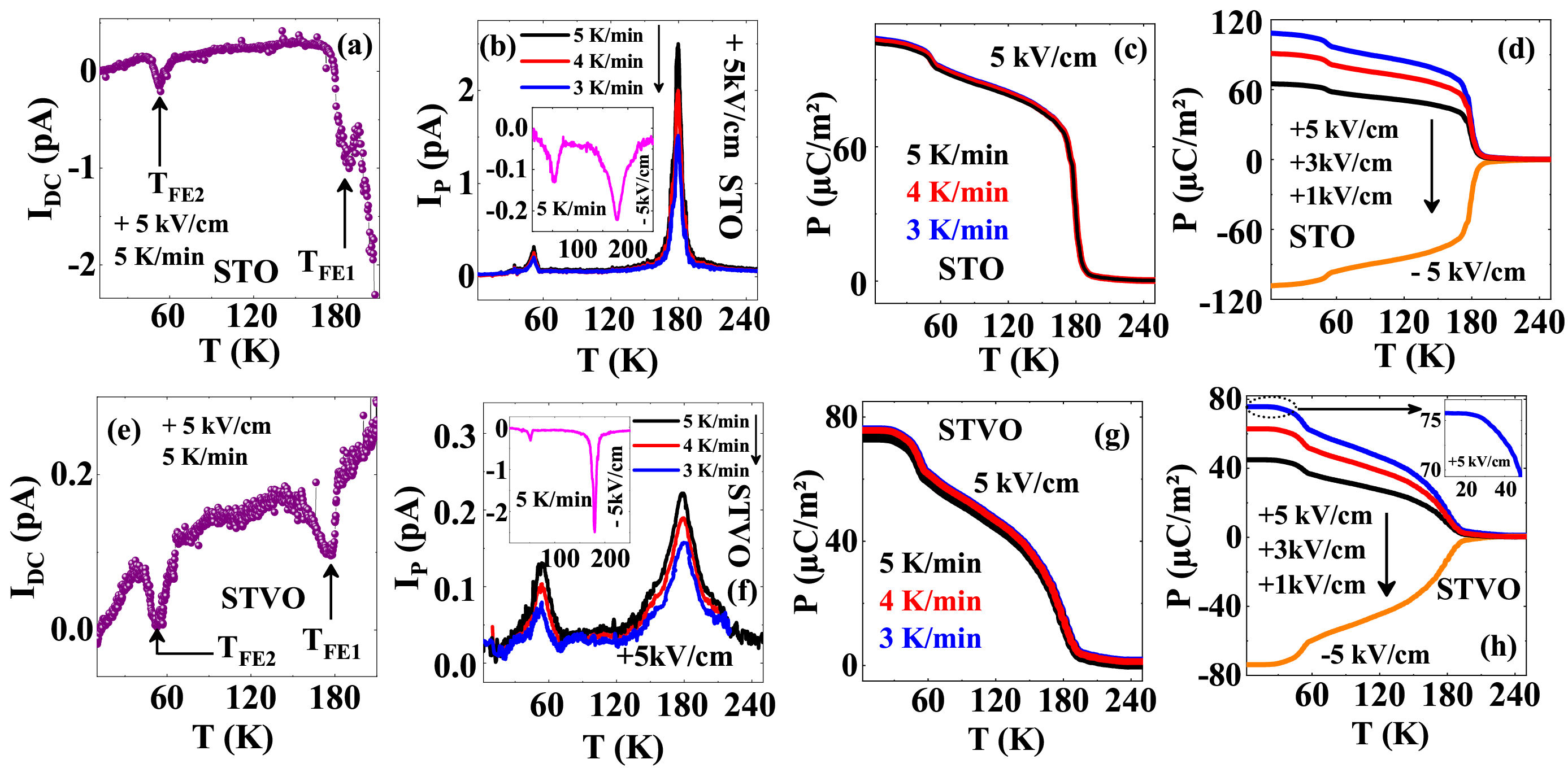}
\caption {$T$ variations of $I_{DC}$ for (a) STO and (e) STVO, $I_P$ for different thermal sweep rates for (b) STO and (f) STVO, polarization ($P$) for different thermal sweep rates for (c) STO and (g) STVO, $P$ at different electric fields for (d) STO and (h) STVO, respectively. Insets of (b) and (f)  show $I_P$ for -5kV/cm poling field. Inset of (h) highlights saturating $P$-value below $T_{FE2}$ for STVO.}
\label{pol1}
\end{figure*}
%%%%%%%%%%%%%%%%%

%The emergence of an anomaly and a change in the slope of dM/dT around 180 K, as well as an anomaly around 52 K for both STVO and STO, are significant. These ferroelectric transitions, observed at 52 K and 180 K, as discussed elsewhere, indicate magnetoelectric coupling. Additionally, a sharp minimum is observed at 272 K for STVO, likely due to short-range magnetic order attributed to the 5% V doping. The probable short-range ordering in STVO requires further confirmation using other experimental methods.

Temperature-dependent magnetization ($M$) recorded in zero-field-cooled (ZFC) mode for STO and STVO is illustrated in Figs. \ref{magnetic}(a) and \ref{magnetic}(c), respectively. In the ZFC mode, samples were cooled in zero-field, and magnetization was recorded in the warming mode with $H$ = 1 kOe. Broad maxima in $M(T)$ are observed around 150 K for both samples, with a sharp rise observed below $\sim$ 35 K, consistent with a previous report on STO \cite{sing1}. The broad maximum in STO was attributed to the crystal field effect of Sm$^{3+}$ \cite{sing1}. For STVO, a new broadened anomaly appears around 270 K, as highlighted in the inset of Fig. \ref{magnetic}(c), which emerges due to 5\% V doping. Plots of $dM/dT$ with $T$ are depicted in Figs. \ref{magnetic}(b) and \ref{magnetic}(d), respectively, highlighting prominent characteristic features by the arrows for STO and STVO. The change of slope in $dM/dT(T)$ around 180 K and an anomaly around 52 K for both STVO and STO, are significant. Because these signatures coincide with the ferroelectric transitions at 52 and 180 K, as discussed elsewhere, indicating magnetoelectric coupling. Additionally, a sharp minimum is observed at 272 K for STVO, likely due to short-range magnetic order attributed to the 5 \% V doping. The probable short-range ordering in STVO requires further confirmation using other experimental methods. 
The isothermal magnetization curves of STVO are depicted in Fig. \ref{MH}(a) at selected $T$. Unlike the linear behavior of the magnetization curves for STO \cite{sing1}, non-linearity in the magnetization curves is evident at low $H$ for STVO, as highlighted in the inset of Fig. \ref{MH}(a) at 2 K and \ref{MH}(b) at 150 K. Arrows in the figures indicate magnetic field, beyond which linearity of the magnetization curves is observed, decrease with increasing $T$. This non-linearity may correlate with the emergence of a short-range magnetic order below 270 K for STVO. 

%%%%%%%%%%%%%%%%%%%%%%%%%%%%%%%%%%%%%%%%%%%%%%%
%\subsection{ Dielectric characteristics}
%%%%%%%%%%%%%%%%%%%%%%%%%%%%%%%%%%%%%%%%%%%%%%%
The dielectric permittivity ($\epsilon$) of STO and STVO were recorded from 2 K to 300 K, for example, at a frequency, $f$= 2007 Hz, under zero-field and with $H$ = 50 kOe. Thermal variations of the real component ($\epsilon'$) and the dielectric loss ($\tan\delta$) are depicted for STO and STVO in Figs. \ref{die}(a) and \ref{die}(b), respectively. The $\epsilon'(T)$ and $\tan\delta(T)$ show anomalies at 180 ($T_{FE1}$) and 52 K ($T_{FE2}$) for both the compounds. Observing anomalies in $\epsilon'(T)$ and $\tan\delta(T)$ at the same temperature primarily indicates the onset of spontaneous polar order\cite{muk,kar}. In the presence of a 50 kOe magnetic field, $\epsilon'(T)$ decrease for both compounds, indicating magneto-dielectric (MD) response. The percentages of MD responses, defined as $\epsilon'(H)/\epsilon'(H=0) - 1$ at 2 K are $2.1$ \% and $2.4$ \% for STO and STVO, respectively and are consistent with the previous reports \cite{Shameek2,muk}.

%%%%%%%%%%%%%%%%%%%%%%%%%%%%%%%%%%%%%%%%%%%%%%%
\begin{figure*}
\centering
\includegraphics[width = 2\columnwidth]{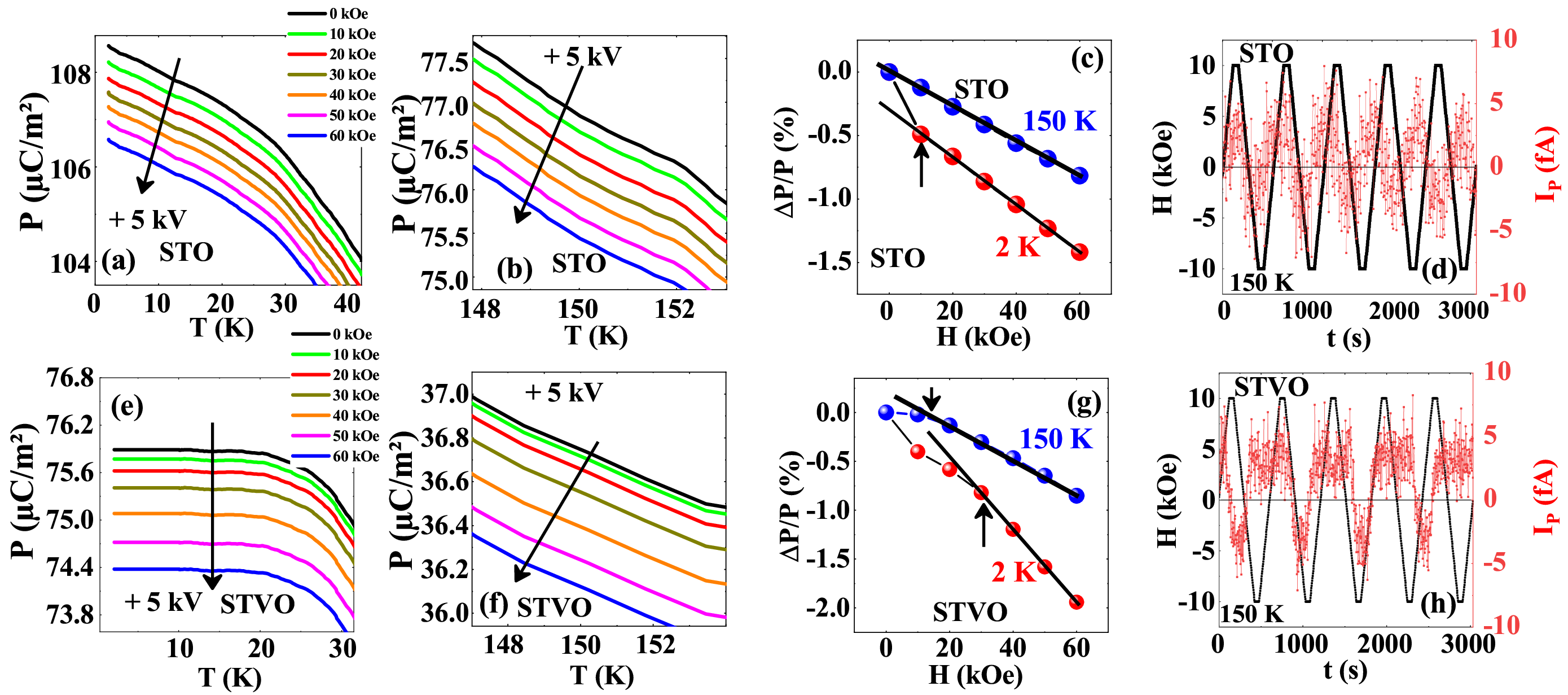}
\caption {$T$ variations of magnified polarization ($P$) under different magnetic fields ($H$) below $T_{FE2}$ for (a) STO and (e) STVO, and below $T_{FE1}$ for (b) STO and (f) STVO, respectively. $H$ variations of $\Delta P/P$ at $2$ and 150 K for (c) STO and (g) STVO, respectively. Time ($t$) dependent oscillations in $H$ and corresponding $I_P$ at 150 K for (d) STO and (h) STVO, respectively.} 
\label{pol2}
\end{figure*}
%%%%%%%%%%%%%%%%%%%%%%%%%%%%%%%%%%%%%%%%%%%%%%%%

%%%%%%%%%%%%%%%%%%%%%%%%%%%%%%%%%%%%%%%%%%%%%

%%%%%%%%%%%%%%%%%%%%%%%%%%%%%%%%%%%%%%%%%%%%%%%
%\subsection{Ferroelectric Ordering and Magneto-electric coupling}
%%%%%%%%%%%%%%%%%%%%%%%%%%%%%%%%%%%%%%%%%%%%%%%

The confirmation of polar order was achieved through various techniques. Upon cooling the sample to its lowest temperature in zero electric field, the bias electric current ($I_{DC}$) was recorded in the presence of a bias electric field of 5 kV/cm, at a rate of 5 K/min in the warming mode \cite{Sundaresan1}. As depicted in Figs. \ref{pol1}(a) and \ref{pol1}(e), anomalies are evident at 52 and 180 K, indicating the onset of long-range polar order for STO and STVO, respectively, consistent with previous reports \cite{Shameek2,Sundaresan1,Terada1}. Further, The pyroelectric current ($I_P$) was recorded under different conditions, such as varied temperature-sweeping rates and different poling electric fields ($E$), including $E$ reversal,  as depicted in Figs. \ref{pol1}(b-d) and \ref{pol1}(f-h). 
The peak positions at $52$ K and $180$ K remain unchanged in these measurements, indicating genuine polar orderings for both the compounds. We observe a reduction in the $I_P$-value with a decrease in temperature-sweeping rate, as depicted in Figs. \ref{pol1}(b) and \ref{pol1}(f) for STO and STVO, respectively. The insets of these figures illustrate $I_P(T)$ at a sweep rate of 5 K/min for a negative poling field of -5 kV/cm. 
Reversing of $I_P(T)$ confirms ferroelectric order. 
Integratiing $I_P(T)$ over time yields electric polarization ($P$) using the equation $P= \int{I_P dt}$. The $P(T)$ curves for STO and STVO are shown in Figs. \ref{pol1} (c) and (g) respectively, for different sweep rates. No significant deviation in $P(T)$ is observed accross different sweep rates,  indicating the absence of extrinsic contributions from thermally stimulated depolarization current (TSDC) \cite{Terada1}. Additionally, we measured $P(T)$ at different poling fields of 5, 3, and 1 kV/cm, as shown in Figs. \ref{pol1}(d) and \ref{pol1}(h). The saturating $P$-values for STO and STVO are $\sim$ 108 and $\sim$ 75 $\mu C/m^2$ respectively, comparable to values for promising multiferroics. \cite{jkd1,muk,Pal1}. While the transition temperatures $T_{FE1}$ and $T_{FE2}$ remain unchanged, a decrease in $P(T)$ is observed with V doping in STVO, consistent with prior findings regarding the effects of doping \cite{Chatterjee,Ramana}. The changes in $P(T)$ at $T_{FE1}$ smears out due to V doping for STVO. In contrast to the behavior observed around $T_{FE1}$, the increase in $P(T)$ at $T_{FE2}$ is more pronounced and abrupt for STVO compared to STO. Although a slight upward trend in $P(T)$ is observed below $T_{FE2}$ down to the lowest recorded temperature for STO, $P(T)$ stabilizes and remains almost constant below $\sim$ 20 K for STVO, as shown in the inset of Fig. \ref{pol1}(h). 

\begin{figure}
\centering
\includegraphics[width =\columnwidth]{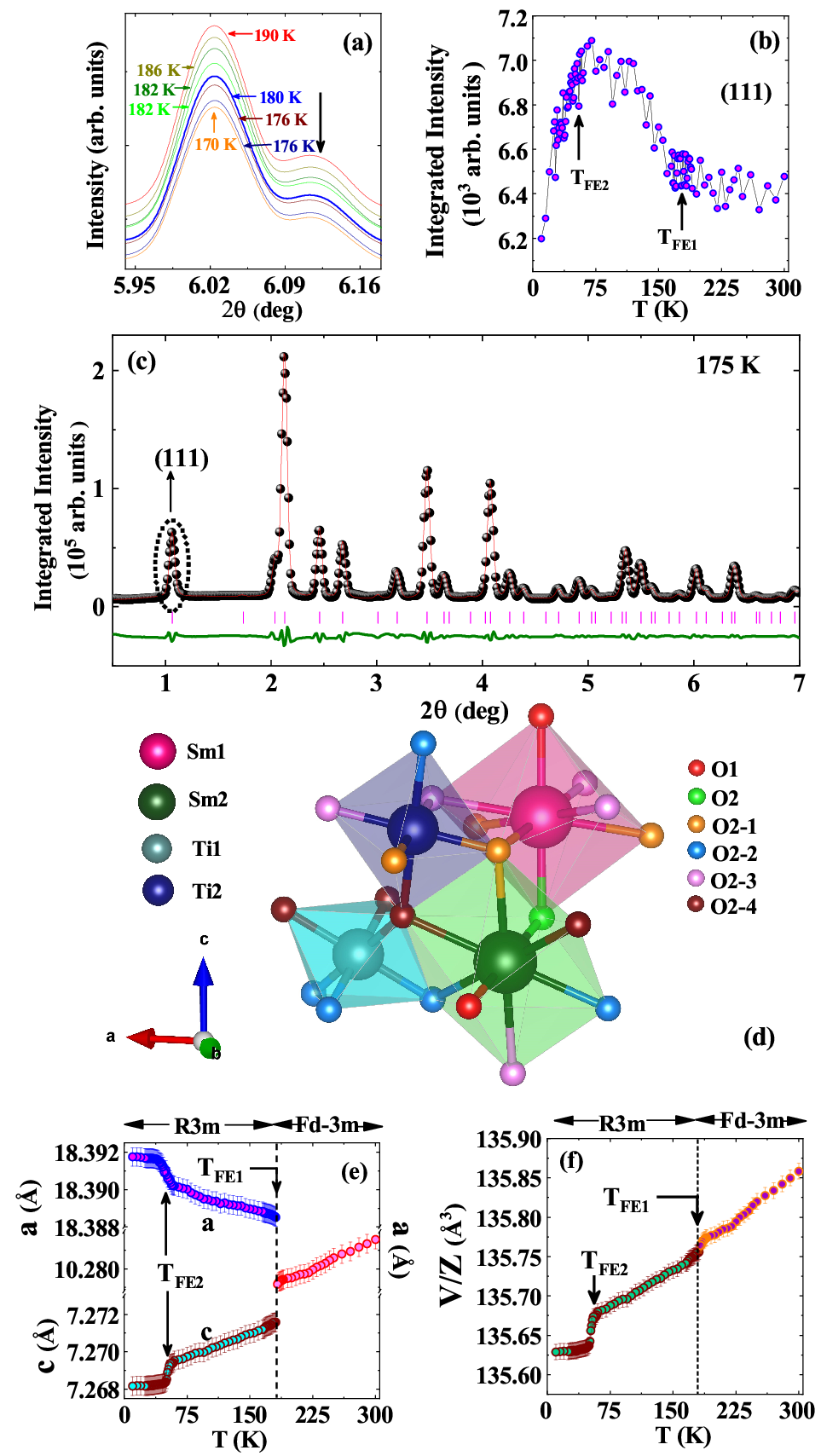}
 \caption {(a) Selected diffraction peaks at different $T$ around $T_{FE1}$. (b) $T$ variation of the integrated intensity of (111) peak. (c) Rietveld refinement of synchrotron diffraction pattern at 175 K. The (111) peak is highlighted by a circle. (d) Connecting Sm-polyhedra and Ti-octahedra below $T_{FE1}$. $T$ variations of lattice parameters (e) $a$ and $c$, and (f) unit cell volume ($V/Z$) for STO, where $Z$ is the number of formula unit within the unit cell.}
\label{sync}
\end{figure}

To investigate the manifestation of ME coupling, the polarization was recorded under various applied magnetic field conditions. The $P(T)$ values at selected $H$ are highlighted for STO and STVO in Figs. \ref{pol2}(a) and \ref{pol2}(e), respectively in the low-$T$ region below $T_{FE2}$, with a poling field of 5 kV/cm. Correspondingly, Figs. \ref{pol2}(b) and \ref{pol2}(f) illustrate the same for the high-temperature range below $T_{FE1}$. The percentage changes in polarization ($\Delta P/P$), calculated as $[P(H) - P(H=0)]/P(H=0)\times$ 100, are plotted against $H$ at 2 K (representative temperature below $T_{FE2}$) and 150 K (representative temperature below $T_{FE1}$) in Figs. \ref{pol2}(c) and \ref{pol2}(g) for STO and STVO, respectively. It is observed that the decrease in $\Delta P/P$ at 2 K surpasses that at 150 K, indicating stronger ME coupling below $T_{FE2}$ for both the cases. A linear decrease of the $P-H$ curves is noted in the high-$H$ region for both STO and STVO. 
The curve remains linear at 150 K and extends into linearity for STO above above 10 kOe. Similarly, for STVO, the $P-H$ curve exhibits linearity above $\sim$ 12 kOe at 150 K and $\sim$ 25 kOe 2 K. Since the initial observation of the linear ME effect in Cr$_2$O$_3$ in the 1960s \cite{Dzyalo,Astrov}, the pursuit of single-phase ME materials demonstrating linear ME coupling has been highly sought after. Recently, linear ME coupling has been identified in LiFePO$_4$ and LiNi$_{0.8}$Fe$_{0.2}$P$_4$ \cite{Fogh}, Na$_2$Co$_2$TeO$_6$ \cite{Shameek2}, Co$_3$V$_2$O$_8$ \cite{muk}, and Gd$_2$MnFeO$_6$ \cite{hati}. Additionally, it is noted that the values of $\Delta P/P$ are $\sim$ 1.5\% and $\sim$ 2\% under 60 kOe and at 2 K for STO and STVO, respectively. An  enhancement of ME is observed due to V doping, albeit with a considerable reduction in the P-value. 

\begin{figure}
\centering
\includegraphics[width =0.7\columnwidth]{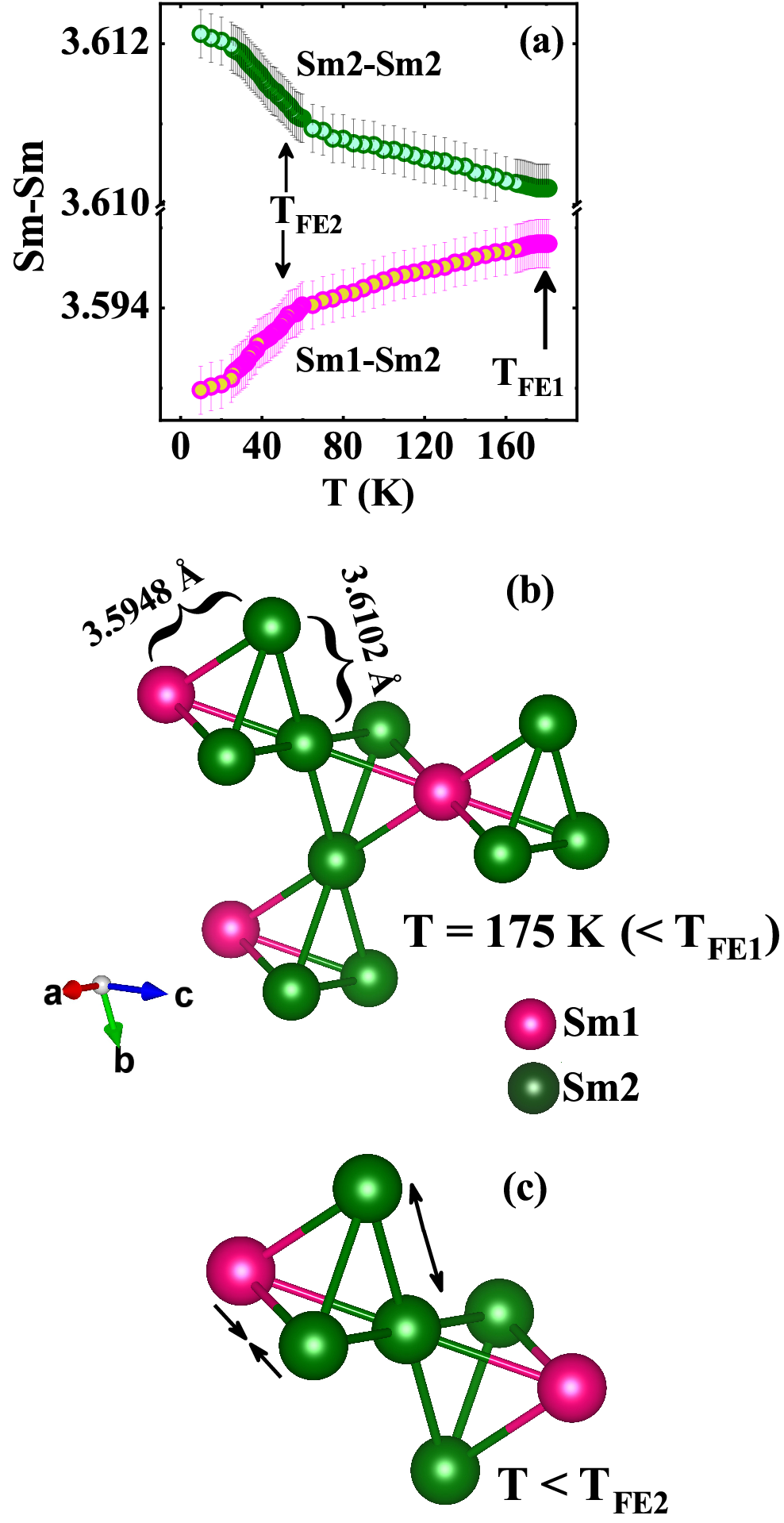}
 \caption {(a) Sm2$-$Sm2 and Sm1$-$Sm2 distances with $T$ below $T_{FE1}$. Schematic representations of distortion of pyrochlore structure below (b) $T_{FE1}$ and (c) $T_{FE2}$.}
\label{distortion}
\end{figure}

The time-dependent behavior of ME was additionally monitored at 150 K for both STO and STVO. In this measurement, sample was initially polarized with an electric field of 5 kV/cm down to 2 K. Subsequently, the electric field was withdrawn, and the sample was properly short-circuited. The sample temperature was then raised to 150 K. The magnetic field was varied between $\pm$ 10 kOe with a sweep rate of 100 Oe/s, while simultaneously recording the $I_P$ over time ($t$), as depicted in Figs. \ref{pol2}(d) and \ref{pol2}(h). 
For STO, the $I_P(t)$ oscillates approximately between $\pm$ 5 fA, exhibiting a phase shift compared to the oscillation of the magnetic field. The value of $I_P(t)$ is small, but it is reproducible and beyond the current measurement resolution of the electrometer. Conversely, the nature of $I_P(t)$ for STVO differs from that of STO. Unlike the regular oscillation $I_P(t)$ for STO, the positive cycle of $I_P(t)$ remains nearly flat and extends over a longer time period compared to sharp change in the negative cycle of $I_P(t)$.

%%%%%%%%%%%%%%%%%%%%%%%%%%%%%%%%%%%%%%%%%%%%%%%%
%\subsection{Low-T synchrotron Diffraction}
%%%%%%%%%%%%%%%%%%%%%%%%%%%%%%%%%%%%%%%%%%%%%%%%

The low-temperature synchrotron diffraction experiments were conducted within the temperature range of 10-300 K for STO. In Figure \ref{sync}(a), a narrow 2$\theta$ region of the diffraction pattern is highlighted around $T_{FE1}$ at selected $T$, with a 4 K interval. While changes in the intensity of the prominent peak are not readily discernible, subtle variations in the intensity of the smaller peak, indicated by an arrow, are faintly observed, with a notable increase in peak height around $T_{FE1}$ at 180 K. 
The integrated intensity of the (111) peak, highlighted in Fig. \ref{sync}(c), is shown in Fig. \ref{sync}(b). The plot indicates a consistent intensity from higher temperatures until $T_{FE1}$, below which it begins to rise quite sharply, reaching a peak around 75 K before declining below $T_{FE2}$. The observed changes in integrated intensity around $T_{FE1}$ resemble those observed in multiferroics, suggesting potential structural transitions, including isostructural changes \cite{Dey1,Dey2} and transitions to the lower symmetries \cite{Indra1,Indra2,jkd1,muk}. No additional diffraction peaks were detected below $T_{FE1}$. However, refinement of the diffraction pattern below $T_{FE1}$ using the room temperature $Fd\bar{3}m$ space group yielded unsatisfactory results, implying a probable structural transition.

%%%%%%%%%%%%%%%%%%%%%%%%%%%%%%%%%%%%%%%%%%%%%%%%

We utilize the open-source web software ISODISTORT \cite{Campbell} to discern potential structures at low temperatures. Our analysis reveals that $R3m$ (no. 227) represents a polar structure, exhibiting the highest symmetry among the various structures considered. This finding aligns with the rhombohedral distortion-driven structural transition to $R3m$ structure from $Fd\bar{3}m$, which was associated with the observed FE order in ZnFe$_2$O$_4$ \cite{jkd2}. 
%We tried to refine the data at low temperatures with this space group and found satisfactory fitting and good fitting parameters, suggesting a better fit than with the $Fd-3m$ (no. 166) structure. 
A satisfactory Rietveld refinement of the diffraction pattern at 175 K, employing the $R3m$ space group, is presented in Fig. \ref{sync}(c). The various Wyckoff positions are as follows: Sm1(0,0,0), Sm2 (0.50,0,0), Ti1(0,0,0), Ti2(0.833333,0.166667,0.166667), O1(0, 0, 0.625), O2(0, 0, 0.375), O2-1(0.532000, 0.468000, 0.141000), O2-2(0.801330, 0.198670, 0.275670), O2-3(0.134670, 0.865330, 0.192330), O2-4(0.865330, 0.134670, 0.057670). This refinement identifies two Sm, two Ti, and six oxygen atoms within the structure. The connectivity of Sm-polyhedra and Ti-octahedra formed by oxygen atoms is illustrated in Fig. \ref{sync}(d). The thermal variations of lattice parameters, $a$ and $c$, and unit cell volume ($V/Z$), as obtained from the refinement, are depicted in Figs. \ref{sync}(e) and \ref{sync}(f), respectively, where $Z$ is the number of formula unit within the unit cell. Besides the structural changes at $T_{FE1}$, the lattice parameters and the volume show another iso-structural transition near T$_{FE2}$.

%%%%%%%%%%%%%%%%%%%%%%%%%%%%%%%%%%%%%
%\section{Discussions and Conclusion}
%%%%%%%%%%%%%%%%%%%%%%%%%%%%%%%%%%%%%%%%%%%%%%%%

%Evidence supporting the release of magnetic frustration has been observed through an iso-structural transition in the widely studied CaBaCo4O7 [52, 53], resulting in the occurrence of ferroelectric order. Additionally, in ZnFe2O4, a structural transition from a cubic structure to a lower symmetry R3m structure is observed to release magnetic frustration and facilitate the occurrence of ferroelectric order. The release of geometric magnetic frustration, attributed to the antiferromagnetic exchange-coupled pyrochlore structure, strongly correlates with this structural transition [50].

The pyrochlore structure in the $R3m$ phase comprises two Sm atoms, denoted as Sm1 and Sm2. Figure \ref{distortion}(a) shows temperature variations of the Sm2$-$Sm2 and Sm1$-$Sm2 distances below $T_{FE1}$. As temperature decreases, the Sm2$-$Sm2 distance increases, while the Sm1$-$Sm2 distance decreases. 
Figure \ref{distortion}(b) demonstrates that the direct distance between Sm1 and Sm2 is smaller than the Sm2$-$Sm2 distance, indicating a distortion in the pyrochlore structure.   
It is significant to note that the equidistant pyrochlore structure formed by Sm atoms at high temperatures, exhibiting the $Fd\bar{3}m$ structure as depicted in Fig. \ref{structure}, leads to significant magnetic frustration, which becomes distorted following the structural transition at $T_{FE1}$. We propose that the release of magnetic frustration is a consequence of the structural transition and highlights its pivotal role in the occurrence of ferroelectric order.
Evidence supporting the release of magnetic frustration has been observed through an iso-structural transition in the widely studied CaBaCo$_4$O$_7$ \cite{Caignaert,kde1}, resulting in the ferroelectric order. Additionally, in ZnFe$_2$O$_4$, a structural transition from a cubic structure to a lower symmetry of $R3m$ structure is observed to release magnetic frustration and facilitates the occurrence of ferroelectric order. The release of geometric magnetic frustration, attributed to the antiferromagnetic exchange-coupled pyrochlore structure, strongly correlates with this structural transition \cite{jkd2}. 
Furthermore, it is noteworthy that the structural change at $T_{FE1}$ in STO bears a resemblance to the well-studied multiferroic BiFeO$_3$, where the $R3m$ structure induces a rhombohedral distortion of the cubic high-temperature perovskite structure formed by the Bi atoms. This structural distortion  is closely associated with the emergence of ferroelectric order in BiFeO$_3$ \cite{Kubel,Park}.    
The emergence of ferroelectric order in the presence of magnetic frustration adds a compelling layer of complexity to these materials, rendering them highly intriguing subjects for further scientific exploration. In Fig. \ref{distortion}(a), the significant changes in slope of Sm$-$Sm distances are observed below $T_{FE2}$, indicating another distortion in the pyrochlore structure. Figure \ref{distortion}(c) shows nature of further distortion in the pyrochlore structure below $T_{FE2}$. These results demonstrate successive distortions in the pyrochlore structure, corresponding to the successive ferroelectric transitions.

Our comprehensive investigation into the structural and ferroelectric characteristics of frustrated pyrochlore compounds, particularly focusing on both the pristine and V-doped STO, has provided rich insights into the interplay between magnetic frustration and ferroelectric order.
We have uncovered distinct ferroelectric phases in both compounds, manifesting around temperatures of $\sim$ 182 K ($T_{FE1}$) and $\sim$ 52 K ($T_{FE2}$). Remarkably, these ferroelectric transitions occur in the absence of long-range magnetic ordering, suggesting that the emergence of ferroelectricity is primarily driven by factors beyond conventional magnetic interactions. 
The introduction of V dopants has revealed intriguing effects on the ferroelectric properties, as evidenced by the observed reduction in polarization values. This reduction may be correlated with the structural disorder in STVO caused by V doping. 
Additionally, the identification of a novel short-range magnetic ordering around 272 K exclusively in the V-doped compound underscores the intricate interplay between magnetic dopants and frustrated pyrochlore structure of the system. 
As highlighted in Fig. \ref{structure}, the direct Sm$-$Sm distance in the pyrochlore structure and the V/Ti$-$Sm distance are exactly the same in the $Fd\bar{3}m$ structure, indicating comparable exchange coupling due to V doping. Therefore, V doping introduces a significant perturbation to the frustrated pyrochlore systems, strongly influencing the ME coupling as well as the polarization value. The occurrence of ME coupling without long-range magnetic order promising for developing ME materials, which can be further enhanced by doping with magnetic elements. 
Looking ahead, future research endeavors could delve deeper into elucidating the underlying mechanisms governing ME coupling in frustrated pyrochlore compounds, especially without long-range magnetic order.

%\section{Conclusion}

To conclude, two ferroelectric orders are revealed well above the dipolar-octupolar antiferromagnetic ordering for STO. A structural transition around 182 K to a polar $R3m$ phase from the high-temperature $Fd\bar{3}m$ space group is associated with the occurrence of higher-temperature ferroelectric ordering. The frustrated pyrochlore structure formed by the Sm atoms is correlated with the ferroelectric order, where the release of magnetic frustration drives a distortion in the pyrochlore structure within the $R3m$ space group, suggesting a correlation with the emergence of ferroelectric order. The replacement of non-magnetic Ti by magnetic $V^{4+}$ strongly influences the ferroelectric polarization as well as the ME coupling in STVO. The value of ferroelectric polarization decreases, while the ME coupling is enhanced due to V doping in STO. The emergence of ME coupling without the involvement of long-range magnetic order is unusual and is further enhanced by V doping. This observation indicates a new pathway for tuning ME coupling.

\vspace{0.2in}
\noindent
{\bf Acknowledgment}\\
S.G. would like to thank BRNS, India (No. 58/14/09/2023/11737), SERB, India, for the CRG project (No. CRG/2022/000718), and the Department of Science and Technology, India, for the financial support (Proposal No. ID-IB-2021-07), and Jawaharlal Nehru Centre for Advanced Scientific Research, Bangalore, India for facilitating the experiments at the beamline P21.1, DESY Synchrotron Facility, Hamburg, Germany. S.G. also would like to acknowledge UGC-DAE Consortium for the instrumental support of scientific research (No. CRS/2022-23/01/685).

\end{document}